# Probing the basis set limit for thermochemical contributions of inner-shell correlation: Balance of core-core and core-valence contributions[1]

Nitai Sylvetsky and Jan M. L. Martin

Department of Organic Chemistry, Weizmann Institute of Science, 76100 Reḥovot, Israel
[a)]Corresponding author: gershom@weizmann.ac.il

**Abstract.** The inner-shell correlation contributions to the total atomization energies (TAEs) of the W4-17 computational thermochemistry benchmark have been determined at the CCSD(T) level near the basis set limit using several families of core correlation basis sets, such as aug-cc-pCVnZ (n=3-6), aug-cc-pwCVnZ (n=3-5), and nZaPa-CV (n=3-5). The three families of basis sets agree very well with each other (0.01 kcal/mol RMS) when extrapolating from the two largest available basis sets: however, there are considerable differences in convergence behavior for the smaller basis sets. nZaPa-CV is superior for the core-core term and awCVnZ for the core-valence term. While the aug-cc-pwCV(T+d)Z basis set of Yockel and Wilson is superior to aug-cc-pwCVTZ, further extension of this family proved unproductive. The best compromise between accuracy and computational cost, in the context of high-accuracy computational thermochemistry methods such as W4 theory, is CCSD(T)/awCV{T,Q}Z, where the {T,Q} notation stands for extrapolation from the awCVTZ and awCVQZ basis set pair. For lower-cost calculations, a previously proposed combination of CCSD-F12b/cc-pCVTZ-F12 and CCSD(T)/pwCVTZ(no f) appears to 'give the best bang for the buck'. While core-valence correlation accounts for the lion's share of the inner shell contribution in first-row molecules, for second-row molecules core-core contributions may become important, particularly in systems like $P_4$ and $S_4$ with multiple adjacent second-row atoms. The average absolute core-core/core-valence ratio is 0.08 for the first-row species in W4-17, but 0.47 for the second-row subset.

## INTRODUCTION

In wavefunction ab initio studies, it used to be taken for granted that correlation of inner-shell electrons did not matter greatly — or that if they mattered at all, they did so less than the remaining 1-particle and n-particle truncation errors in the valence correlation energy (see below). However, as quantitative "chemical accuracy" (traditionally defined as ±1 kcal/mol) came in sight, this view needed to be revised.[1,2]

Composite computational thermochemistry schemes such as G3 and G4 theory,[1,3] CBS-QB3,[4–6] the ccCA approach of Wilson and coworkers,[7–9] Weizmann-n theory (Wn),[2,10–13] the explicitly correlated Wn-F12 variants thereof,[14–17] and the FPD (Feller-Peterson-Dixon) approach[18–22] all include an inner-shell correlation component. These all are based on some form of the following decomposition:

$$E = E_{SCF} + \Delta E_{corr,valence} + \Delta E_{inner-shell} + \Delta E_{scalar-rel} + \Delta E_{S-O} + \Delta E_{DBOC} + ZPVE \qquad (1)$$

in which the different terms represent the SCF energy, the valence correlation energy, the inner-shell correlation energy, the scalar relativistic correction, the (first-order) spin-orbit correction, the diagonal Born-Oppenheimer correction, and the (anharmonic or scaled-harmonic) zero-point vibrational energy, respectively. (For G3 and G4, an empirical correction term for residual basis set incompleteness is added, its coefficients fitted to experimental heats of formation.) In G4, CBS-QB3, and ccCA, $\Delta E_{corr,valence}$ is approximated as $\Delta E_{MP2,valence} + \Delta E[CCSD(T)–MP2]/small$, the latter term being evaluated using quite small basis sets. In Wn, $\Delta E_{corr,valence}$ is instead partitioned as

$$\Delta E_{val}[CCSD] + \Delta E_{val}[(T)] + \Delta E_{val}[T_3–(T)] + \Delta E_{val}[(Q)] + \Delta E_{val}[T_4-(Q)] + \Delta E_{val}[T_5] \tag{2}$$

---

[1] In memory of Dieter Cremer (1944-2017)

Each successive term in this equation is obtained or extrapolated from successively smaller basis sets. (Similar components, with some variations, are seen in the older FPD and in the Focal Point Approach.[23–25]) The HEAT method developed by an international consortium around John F. Stanton,[26–28] especially in its most recent version,[27] substantially contains the same contributions, but eschews core-valence partitioning for the CCSD(T) correlation energy. (This somewhat encumbers its applicability to 2nd-row systems.) G4 and CBS-QB3 do not include relativistic corrections at all but attempt to absorb these effects into empirical corrections.

Specifically for W4, it has been shown[29] by comparison to Active Thermochemical Tables[30–36] (AtcT) data, that 3RMSD≤1 kJ/mol accuracy is achievable for total atomization energies; substantially the same should be true for other high-accuracy computational thermochemistry schemes such as HEAT and FPD. What factors limit accuracy in practice? For systems like $O_3$ or $F_2O_2$ with significant nondynamical correlation,[37] the largest uncertainty would be associated with the post-CCSD(T) corrections. On the other hand, for systems like $CCl_4$ — which are dominated by dynamical correlation but exhibit somewhat refractory basis set convergence — that would be the valence CCSD correlation component.[17] The scalar relativistic, spin-orbit, and DBOC terms converge fairly rapidly with the basis set,[12] while the issue of the ZPVE is discussed at length elsewhere.[38] This leaves the inner-shell correlation term, which is the focus of the present paper.

In electron correlation methods that trivially decompose into pair correlation energies, such as MP2, MP3, CISD, and CCSD, each pair can be assigned as valence, core-valence, or core-core, depending on whether neither, one, or both of the two occupied spin-orbitals are in the subvalence shell. (According to Nesbet's theorem, this partitioning can be done more generally, at least for closed-shell cases.[39]) Thus, the overall correlation energy can then also be partitioned into valence, core-valence (CV), and core-core (CC). To the best of our knowledge, Bauschlicher, Langhoff, and Taylor (BLT)[40] were the first to point out that, at least for first-row compounds, the lion's share of inner-shell correlation effects on molecular properties comes from core-valence correlation: the core-core contribution tends to cancel between the molecule and its separated atoms. (This partitioning can readily be extended to multiple subvalence shells, e.g., subvalence-valence (n-1)d correlation in heavy p-block atoms.) It has since been widely accepted in the ab initio community that all chemically relevant effects of inner-shell correlation arise from E(CV), to the extent that "core-valence correlation" has become a synecdoche for all inner-shell correlation effects.

Accounting for inner-shell correlation requires basis sets adapted to the purpose, particularly in terms of radial flexibility in the inner-shell region. Martin and Taylor[41,42] developed the so-called "MTsmall" basis set in an *ad hoc* fashion. It is still used for the core-valence and scalar relativistic contributions in W1 theory.[2,43] Later, Peterson, Dunning, and coworkers more systematically developed core-valence correlation basis sets cc-pCVnZ (n=D,T,Q,5) for the first row,[44] and later for the second row.[45] The cc-pCV6Z basis sets for both rows were published much later.[46] Unlike in first-row elements, the additional (n-1)s and (n-1)p shells in second-row elements mean that CC+CV absolute correlation energies rival or exceed valence correlation energies. Consequently, a strict energy optimization for E(CC+CV) yields basis sets that are less efficient at recovering the chemically important CV term, and hence Peterson and Dunning developed[45] the alternative "core-valence weighted" basis sets, cc-pwCVnZ, in which optimization is instead carried out to a weighted average of E(CC) and E(CV), the weighting heavily biased toward E(CV). Similar basis sets were later developed for heavy p-block elements[47] and for transition metals,[48–51] while additional such basis sets were developed for use with small-core pseudopotentials.[52]

(The case of alkali and alkaline earth metals, where (n-1)p orbitals may actually cross into the valence shell and thus acquire "honorary valence orbital" character, needs special consideration. Core-valence basis sets for these elements were developed by Iron et al.[53] and by Hill et al.[54])

The subvalence correlation energy in second-row compounds can be an order of magnitude larger than in first-row species. Somewhat counterintuitively, early benchmark surveys (e.g., in the framework of W1 theory[2,10]) revealed that this does not necessarily translate into greater numerical importance for reaction energies (e.g., atomization energies).

Quite recently, Ranasinghe et al.[55] published a new family of nZaPa-NR_CV basis sets. These were obtained from the earlier nZaPa-NR valence basis sets[56] by optimizing basis functions for correlation in He-like ions ($B^{3+}$, $C^{4+}$,..) and in Ne-like ions ($Al^{3+}$, $Si^{4+}$,…), then adding some exponents in each angular momentum by interpolating between the valence and core functions. By construction, these basis sets favor core-core correlation.

Very recently, we published the W4-17 thermochemical benchmark,[57] an expanded update of the earlier W4-11 dataset.[29] W4-17 consists of 200 small first-row and second-row molecules, spanning a broad range of bonding situations and degrees of nondynamical correlation. Core-valence contributions in all these had been obtained at the CCSD(T) level[58,59] and extrapolated from aug-cc-pwCVTZ and aug-cc-pwCVQZ basis sets (shorthand notation: awCV{T,Q}Z); this level had previously been found adequate by benchmarking against a fairly small sample.

In the present study, we shall obtain benchmark core-valence correlation contributions for the entire W4-17 dataset at the 1-particle basis set limit, assess the performance of more cost-effective basis methods (both conventional and explicitly correlated) and show that a number of trends in the core-valence contributions can be rationalized in terms of E(CC)+E(CV) decomposition.

For the sake of completeness, we mention a bond additivity model,[60] as well as a study by Nicklass and Peterson[61] in which shown that CC+CV contributions for some 1st-row atoms predicted surprisingly well (at least semiquantitatively) by a core polarization potential (CPP).

Furthermore, Ranasinghe, Petersson and Frisch[62] also developed a semiempirical DFT functional for the estimation of core-valence correlation energies.

## COMPUTATIONAL METHODS

All conventional and explicitly correlated *ab initio* calculations were carried out using the MOLPRO 2015.1 program system[63,64] running on the Faculty of Chemistry cluster at the Weizmann Institute of Science.

The augmented core correlation consistent (ACVnZ) basis sets by Dunning and coworkers,[44–46] the nonrelativistic n-tuple-ζ augmented polarization and core-valence augmented (nZaPa-NR_CV) basis sets by Ranasinghe et al.,[55] and the augmented core-valence weighted (awCVnZ) sets by Peterson and coworkers[45] were used for the conventional coupled-cluster single point energy calculations included in our work.

Recently (e.g.,[14,17]), explicitly correlated methods (see[65,66] for recent reviews) have proven to be very valuable for thermochemical applications, on account of their much faster basis set convergence for the valence correlation energy. In the present work, we will assess their performance for core-valence contributions as well. To this end, CCSD-F12b calculations[67,68] were performed using the cc-pCVnZ-F12 basis sets (n = D, T, Q).[46] These latter basis sets will be denoted CVnZ-F12 from now on. Optimal values for the geminal Slater exponents (β) used in conjunction with the CVnZ-F12 basis sets were taken from Ref[46] as well. "CABS corrections"[68,69] were employed throughout for the calculated SCF components, and 3C(Fix) ansatz[70] and the CCSD-F12b approximation[67,68] were used throughout.

Explicitly correlated CCSD-F12b calculations in practical implementations require not only the specification of an orbital basis set and a geminal exponent, but also of three auxiliary basis sets, denoted here by their acronyms in the MOLPRO program system.[63] In the present work, the JKfit fitting basis sets of Weigend[71] were employed for the density fitting in the Hartree–Fock calculations, while the MP2fit set of Hättig and co-workers[72] was employed for the density fitting of the remaining two-electron integrals in the CCSD-F12b calculations. The RI approximation was applied using the OptRI or "CABS" (or CABS, complementary auxiliary basis sets[73]) of Yousaf and Peterson.[74]

Basis set extrapolations were carried out using the two-point formula:

$$E_\infty = E(L) - [E(L) - E(L-1)] / \left[ \left(\frac{L}{L-1}\right)^\alpha - 1 \right] \quad (3)$$

where L is the angular momentum in the basis set and α an exponent specific to the level of theory and basis set pair. [Unless noted otherwise, basis set extrapolation exponents (α) were taken from the compilation in Table 2 of Ref[75].]

The inner-shell correlation contribution as reported and discussed here decomposes as follows:

$$E_{corr}[CCSD, \text{inner-shell}]_{\text{inner-shell}} = E_{corr}[CCSD,CC] + E_{corr}[CCSD,CV] + E_{corr}[CCSD,\text{exclusion}] + E_{corr}[(T)] \quad (4)$$

where E[CCSD,CC] and E[CCSD,CV] are the sums of the core-core and core-valence CCSD pair correlation energies, respectively, as reported by MOLPRO 2015.1, and the "exclusion energy"[55,76] corresponds to the sum of valence CCSD pair correlation energies in the inner-shell calculation minus the corresponding sum in a valence-only CCSD calculation (which equals the valence CCSD correlation energy).

$$E_{corr}[CCSD, \text{exclusion}] = E_{corr}[CCSD_{\text{inner-shell}}, VV] - E_{corr}[CCSD_{\text{valence}}] \quad (5)$$

In plain English, this term corresponds to the reduction in the valence correlation energy due to the fact that valence electrons now have to 'compete' with the inner-shell for access to the virtual orbital space.

# RESULTS AND DISCUSSION

## Basis set convergence of CC and CV

Table 1 contains a summary of the basis set convergence behavior of core-core and core-valence components for the three main sequences of basis sets considered here, namely, aCVnZ (n=3–6), awCVnZ (n=3-5), and nZaPa-CV (n=3-5). We were able to perform aCV6Z calculations for a large subset of W4-17, but had to abandon these calculations for the remaining molecules for reasons of near-linear dependence (e.g., $C_2H_6$) or computational demands (e.g., $C_2Cl_6$). Where available, we use CCSD(T)/aCV{5,6}Z as the primary reference.

**TABLE 1.** RMSD (kcal/mol) for the W4-17 TAE Benchmark of Core-core and Core-valence Components of the CCSD Part

| n= | T | Q | 5 | 6 | {T,Q} | {4,5} | {5,6} |
|---|---|---|---|---|---|---|---|
| Core-core w.r.t. ACV{5,6}Z | | | | | | | |
| awCVnZ | 0.120 | 0.058 | 0.028 | | 0.023 | 0.004 | |
| ACVnZ | 0.097 | 0.027 | 0.012 | 0.007 | 0.025 | 0.004 | REF |
| nZaPa-CV | 0.010 | 0.006 | 0.004 | | 0.005 | 0.003 | |
| Core-valence w.r.t. ACV{5,6}Z | | | | | | | |
| awCVnZ | 0.221 | 0.084 | 0.040 | | 0.021 | 0.009 | |
| ACVnZ | 0.264 | 0.112 | 0.055 | 0.030 | 0.054 | 0.016 | REF |
| nZaPa-CV | 0.248 | 0.077 | 0.035 | | 0.051 | 0.015 | |
| Core-core w.r.t. awCV{Q,5}Z | | | | | | | |
| awCVnZ | 0.140 | 0.063 | 0.030 | | 0.020 | REF | |
| ACVnZ | 0.100 | 0.029 | 0.013 | 0.007 | 0.026 | 0.006 | 0.004 |
| nZaPa-CV | 0.016 | 0.010 | 0.007 | | 0.009 | 0.006 | |
| Core-valence w.r.t. awCV{Q,5}Z | | | | | | | |
| awCVnZ | 0.265 | 0.103 | 0.050 | | 0.028 | REF | |
| ACVnZ | 0.267 | 0.117 | 0.057 | 0.030 | 0.062 | 0.014 | 0.009 |
| nZaPa-CV | 0.326 | 0.104 | 0.040 | | 0.059 | 0.009 | |

REF denotes the reference level in this and following tables

Considering the way that the nZaPa-CV basis sets are constructed, it is not surprising that for the core-core term, their convergence is by far the most rapid of all three sequences: 3ZaPa-CV even outperforms aCV5Z, in fact. For the converse reason, it is not surprising that awCVnZ, being explicitly optimized for core-valence, displays the slowest convergence of the three for the core-core term. Yet when extrapolating from n={4,5}, all three sequences are within 0.003 to 0.004 kcal/mol RMS from our best available reference. When going back to n={T,Q}, performance of aCV{T,Q}Z and awCV{T,Q}Z is actually quite similar (0.025 and 0.023 kcal/mol RMS, respectively). {3,4}ZaPa-CV yields just 0.005 kcal/mol, but it should be kept in mind that the nZaPa-CV basis sets are considerably larger (Table 2) than the corresponding awCVnZ and aCVnZ basis sets — especially for the second row, where nZaPa-CV has sizes comparable to awCV(n+1)Z.

**TABLE 2.** Numbers of Basis Functions in Different Basis Sets

| | 1st-row | | | | |
|---|---|---|---|---|---|
| | D | T | Q | 5 | 6 |
| CVnZ-F12 | 34 | 62 | 96 | - | - |
| awCVnZ | - | 59 | 109 | 181 | - |
| ACVnZ | - | 59 | 109 | 181 | 279 |
| nZaPa-CV | - | 70 | 132 | 213 | - |
| | 2nd-row | | | | |
| | D | T | Q | 5 | 6 |
| CVnZ-F12 | 48 | 78 | 112 | - | - |
| awCVnZ | - | 75 | 134 | 217 | - |
| ACVnZ | - | 75 | 134 | 217 | 324 |
| nZaPa-CV | - | 118 | 209 | 326 | - |

For core-valence, it is not surprising that awCVnZ converges more rapidly than ACVnZ or nZaPa-CV — but when extrapolating {4,5}, all three sequences come closer than 0.02 kcal/mol RMS. Out of the three, awCV{Q,5}Z comes closest at 0.009 kcal/mol RMS — and since we were able to cover all 200 systems at the CCSD(T)/awCV{Q,5}Z level, we are using this as our reference throughout the remainder of the paper.

As noted earlier,[55] the exclusion energy[55,76] converges rapidly with the basis set, and it is not a factor in the decision between different basis set sequences.

**TABLE 3.** RMSD (kcal/mol) for the W4-17 TAE Benchmark of Connected Triple Excitations, (T) Component of Inner-Shell Contributions

| | D | T | Q | 5 | 6 | {T,Q} | {Q,5} |
|---|---|---|---|---|---|---|---|
| awCVnZ | | 0.062 | 0.022 | 0.010 | | 0.008 | (REF) |
| ACVnZ | | 0.077 | 0.027 | 0.011 | 0.006 | 0.011 | 0.002 |
| ACVn+dZ | | 0.074 | 0.032 | | | | |
| nZaPa-CV | | 0.036 | 0.014 | 0.006 | | 0.004 | 0.001 |
| CVnZ-F12 | 0.172 | 0.054 | 0.034 | | | | |

The (T) in the CCSD(T)-F12b/CVnZ-F12 calculations are unscaled

As for the contribution of (T), in Table 3 it can be seen that aCV{5,6}Z, awCV{Q,5}Z, and {4,5}ZaPa-CV are functionally equivalent in quality, agreeing to 0.001 kcal/mol. We hence can safely use awCV{Q,5}Z as the reference, which we have available for all 200 systems. While 3ZaPa-CV at first sight appears to gain one "n step" over awCVnZ and ACVnZ, this is offset by the extra cost nearly equivalent to one such n-step. At any rate, both awCV{T,Q}Z and ACV{T,Q}Z are already within 0.01 kcal/mol of the reference, and hence this is not the convergence-limiting factor.

Hence, considering that awCV{Q,5}Z is very close to ACV{5,6}Z for core-core and (T), and (by construction) superior over ACV{Q,5}Z for core-valence, we have chosen CCSD(T)/awCV{Q,5}Z as the secondary reference, which is available for the entire set.

Let us now consider CCSD(T) inner-shell contributions in total. As can be seen in Table 4, the RMS deviation between CCSD(T)/awCV{Q,5}Z and the partial CCSD(T)/ACV{5,6}Z inner-shell corrections is only 0.010 kcal/mol: for the CCSD(T)/ACV{Q,5}Z and {4,5}ZaPa-CV data, the RMSD is statistically indistinguishable at 0.011 and 0.010 kcal/mol, respectively. When 2nd-row systems are being considered in isolation (lower pane of Table 4), it appears that the awCVnZ and nZaPa-CV sequences converge more rapidly than ACVnZ: keeping in mind the greater size of nZaPa-CV, awCVnZ emerges as the winner. This is indeed not surprising considering the greater importance of (2s,2p) correlation for these systems.

A remark about the accuracy of our reference data is appropriate. As can be seen in Table 1, for the core-core contribution, the extrapolation needs to cover less than 0.01 kcal/mol from our largest basis sets, and our two extrapolated limits agree to within 0.004 kcal/mol. An even better level of convergence was established for the (T) component (Table 3), where awCV{Q,5}Z, ACV{Q,5}Z, and {4,5}ZaPa-CV all agree to 0.002 kcal/mol RMS. This leaves the core-valence component as the accuracy-determining factor: extrapolation covers 0.03 kcal/mol between our largest basis set and the extrapolated limits. However, this still only translates into a difference of 0.009 kcal/mol between the ACV{5,6}Z, awCV{Q,5}Z, and {4,5}ZaPa values. Considering also the statistics for the totals in Table 4, we believe that the RMS difference of 0.01 kcal/mol is a realistic estimate for the average uncertainty in our reference values.

**TABLE 4.** RMSD (kcal/mol) for the W4-17 TAE Benchmark of Complete CCSD(T) Inner-shell Contributions

| | \multicolumn{7}{c}{Full dataset} | | | | | | |
|---|---|---|---|---|---|---|---|
| | T | Q | 5 | 6 | {T,Q} | {Q,5} | {5,6} |
| awCVnZ | 0.219 | 0.084 | 0.040 | | 0.027 | (REF) | |
| ACVnZ | 0.345 | 0.142 | 0.054 | 0.028 | 0.067 | 0.011 | 0.010 |
| ACVn+dZ | 0.347 | 0.143 | | | 0.065 | | |
| awCVn+dZ | 0.208 | 0.082 | | | 0.034 | | |
| nZaPa-CV | 0.333 | 0.103 | 0.040 | | 0.074 | 0.010 | |

| | \multicolumn{7}{c}{2nd-row Cases Only} | | | | | | |
|---|---|---|---|---|---|---|---|
| | T | Q | 5 | 6 | {T,Q} | {Q,5} | {5,6} |
| awCVnZ | 0.262 | 0.099 | 0.048 | | 0.039 | (REF) | |
| ACVnZ | 0.356 | 0.135 | 0.067 | 0.033 | 0.083 | 0.023 | 0.017 |
| ACVn+dZ | 0.283 | 0.123 | | | 0.073 | | |
| awCVn+dZ | 0.185 | 0.088 | | | 0.048 | | |
| nZaPa-CV | 0.269 | 0.093 | 0.041 | | 0.048 | 0.013 | |

Yockel and Wilson noted[77] that the ACVTZ basis set, which is used in the ccCA approach[7–9] for computing the core-valence correction, was inadequate for (pseudo)hypervalent systems of the type where tight d functions are essential in the *valence* basis set. They noted that the original cc-pCVTZ basis sets had core-valence functions optimized on top of the regular (aug-)cc-pVTZ basis sets and instead optimized aug-cc-pCV(n+d)Z basis sets where an extra *d* has been added to 2nd-row elements. For the 2nd-row subset of W4-17, ACV(T+d)Z does have a lower RMSD=0.28 kcal/mol compared to 0.36 kcal/mol for ACVTZ, though awCVTZ actually attains marginally better RMSD=0.26 kcal/mol. Between ACVQZ and ACV(Q+d)Z, however, the difference is quite small, 0.135 to 0.123 kcal/mol, both inferior to awCVQZ, 0.099 kcal/mol. We also obtained ACV(5+d)Z results for a subset of the second-row molecules and found differences with regular ACV5Z of 0.001 kcal/mol or less. Again, awCV5Z is superior to either, 0.048 vs. 0.067 kcal/mol.

By way of experiment, we optimized core-valence weighted versions of Yockel and Wilson's basis sets, which we term awCV(n+d)Z. We used the same averaging as in the original cc-pwCVnZ paper, except that we employed CCSD core-core and core-valence energies instead of CISD. (The function optimization module in MOLPRO was employed for this purpose.) For n=5 we were unable to obtain satisfactory convergence, as was the case for awCV6Z when we attempted such optimizations. For n=T,Q we obtained solutions: the basis sets are available in the Electronic Supporting Information to the present paper. As can be seen below, the awCV(T+d)Z basis set, at RMSD=0.185 kcal/mol for the 2$^{nd}$-row subset, is superior to awCVTZ, ACVTZ, and ACV(T+d)Z all alike. For n=Q, awCV(Q+d)Z still has a slight edge over awCVQZ. Upon {T,Q} extrapolation, however, the RMSD of awCV({T,Q}+d)Z is actually slightly poorer than that of awCV{T,Q}Z with either being definitely superior over ACV{T,Q}Z at 0.083 kcal/mol and ACV({T,Q}+d)Z at 0.073 kcal/mol. When we consider the entire 200-molecule sample, then the difference between ACV{T,Q}Z and ACV({T,Q}+d)Z is altogether hard to detect, and awCV({T,Q}+d)Z at RMSD=0.034 kcal/mol is probably statistically equivalent, within the remaining uncertainty in the reference data, to the RMSD=0.027 kcal/mol of awCV{T,Q}Z. We conclude that the only way to obtain better core-valence data than those used in W4 and W4-F12 theory would be to proceed to awCV{Q,5}Z basis sets.

Now what about CVnZ-F12? As can be seen in Table 5, for a given cardinal number (n in ACVnZ, awCVnZ, CVnZ-F12, etc), the CCSD-F12b results are about one "n step" closer to the CCSD limit than available alternatives. For example, the RMSD for CVTZ-F12 (0.087 kcal/mol) is similar to that obtained by awCVQZ (0.082 kcal/mol); Similarly, CVQZ-F12 (0.031 kcal/mol RMSD) outperforms awCV5Z (0.039 kcal/mol RMSD). CV{D,T}Z-F12 extrapolation reaches as low as 0.05 kcal/mol, while CV{T,Q}Z-F12 does not offer an advantage over awCV{T,Q}Z. Excluding the latter case, the CVnZ-F12 basis sets offer an accuracy similar to that obtained by much larger conventional basis sets (see Table 4): The CVQZ-F12 set, for instance, is about half the size of awCV5Z or ACV5Z for both 1$^{st}$- and 2$^{nd}$-row elements. Considering the asymptotic $O(N^4)$ (with N the basis set size) CPU time scaling of CCSD and CCSD(T), this translates into about a factor of 16 in CPU time (and mass storage) requirements.

When it comes to the (T) component, however, the CVnZ-F12 do not perform satisfactorily: CVTZ-F12 and CVQZ-F12 display RMS deviations of 0.054 and 0.034 kcal/mol, respectively, while awCVQZ performs well, with just 0.022 kcal/mol RMSD. Furthermore, awCV{T,Q}Z is getting very close to the reference values, with just 0.008 kcal/mol RMSD, and as such is a much better option. Marchetti-Werner scaling[78] of the core-valence (T) contributions, if used in unmodified form, creates too many size-consistency issues here: they could in principle be eliminated by arbitrarily choosing the $E_{corr}$[MP2-F12]/$E_{corr}$[MP2] ratio for the molecule including core-valence correlation to be the source of the scaling factor, which is somewhat clumsy in practice. The use[79] of a single constant scaling factor specific to the basis set, (Ts), was discussed in Ref.[57] but still entails a large basis set for (T) and hence obviates many of the advantages of an F12 approach. In the context of W1-F12 and W2-F12, Karton et al.[14,57] advocate the use of the small cc-pwCVTZ(no f) basis set in conjunction with a fixed scaling factor or 1.1; for a subset of W4-17, it was found in Ref.[57] that in conjunction with CCSD-F12b/CV{D,T}Z-F12, this yields an RMSD of only 0.07 kcal/mol. In the present work, combining this scaled small basis (T) with CCSD-F12b/CVTZ-F12 yields RMSD=0.10 kcal/mol, which is reduced to just 0.06 kcal/mol with CV{D,T}Z-F12. While this is smaller than other error sources in W1-F12 and W2-F12, for higher-accuracy methods CCSD(T)/awCV{T,Q}Z still offers the best accuracy-cost compromise.

**TABLE 5.** Comparison of CCSD and CCSD-F12 for the Inner-Shell Contribution to TAEs (W4-17 Benchmark, RMSD in kcal/mol)

|            | D     | T     | Q     | 5     | 6     | {D,T} | {T,Q} | {Q,5} | {5,6} |
|------------|-------|-------|-------|-------|-------|-------|-------|-------|-------|
| awCVnZ     |       | 0.215 | 0.082 | 0.039 |       |       | 0.025 | (REF) |       |
| ACVnZ      |       | 0.337 | 0.135 | 0.053 | 0.027 |       | 0.072 | 0.010 | 0.010 |
| ACVn+dZ    |       | 0.324 | 0.134 |       |       |       | 0.075 |       |       |
| awCVn+dZ   |       | 0.200 | 0.090 |       |       |       | 0.053 |       |       |
| nZaPa-CV   |       | 0.302 | 0.092 | 0.035 |       |       | 0.065 | 0.009 |       |
| CVnZ-F12   | 0.274 | 0.087 | 0.031 |       |       | 0.051 | 0.033 |       |       |

# CC and CV decomposition

We are now also in a position to re-assess whether core-valence correlation always outweighs core-core correlation in total atomization energies. A breakdown for some representative examples is given in Table 6.

**TABLE 6.** Breakdown of CCSD(T)/awCV5Z Inner-Shell Contributions to the TAE (kcal/mol) for Some Representative Examples

|  | CCSD | CCSD(T) | (T) | CC | CV | exclusion |
|---|---|---|---|---|---|---|
| n-$C_5H_{12}$ | **5.61** | **5.84** | **0.22** | **0.55** | **4.67** | **0.40** |
| $C_6H_6$ | **6.71** | **7.11** | **0.41** | **0.42** | **5.98** | **0.30** |
| $C_2H_2$ | **2.23** | **2.41** | **0.18** | **0.05** | **2.10** | **0.08** |
| CO | **0.73** | **0.93** | **0.20** | **-0.11** | **0.86** | **-0.02** |
| $C_2F_6$ | **1.66** | **2.16** | **0.50** | **0.35** | **1.26** | **0.04** |
| $C_2Cl_6$ | **1.56** | **2.71** | **1.15** | **-0.64** | **2.95** | **-0.75** |
| $P_4$ | **0.35** | **1.78** | **1.43** | **-0.75** | **2.07** | **-0.96** |
| $S_4(C_{2v})$ | **-0.73** | **0.77** | **1.50** | **-0.27** | **0.90** | **-1.37** |
| $Si_2H_6$ | **-0.10** | **-0.19** | **-0.10** | **-1.98** | **0.89** | **0.99** |
| OCS | **1.26** | **1.69** | **0.43** | **-0.40** | **1.83** | **-0.18** |
| $CS_2$ | **1.00** | **1.66** | **0.66** | **-0.71** | **2.16** | **-0.45** |

For hydrocarbons, and most typical 1st-row organic molecules, we see the following behavior (illustrated in Table 6 using n-pentane and benzene): small attractive CC contribution, sizable attractive CV contribution accounting for the lion's share of the total, small attractive (T) contribution (larger in the aromatic ring than in the alkane). In fact, (T) contributions are found to be attractive for nearly all species, except for some B, Al, and Si hydrides. The final term, "exclusion", results from the reduction in the valence correlation energy when the same virtual orbital space also needs to accommodate excitations from the cores.

When triple or short double bonds enter, we see a minor change: the CC contribution becomes nearly zero or even slightly repulsive. We tentatively attribute this to the bulkier (2s,2p) cores getting 'crowded'. [For the sake of illustration, a table of radial density maxima for core orbitals of the elements, taken from the numerical HF data reported by Mann,[80] is given in the Supporting Information.]

In a molecule like $C_2Cl_6$, we see a significant repulsive CC contribution paired with an attractive CV contribution, and a (T) that is over 40% of the total. Similar phenomena are seen for other such chlorides—the more chlorine atoms packed closely together, the more repulsive CC. In contrast, in $C_2F_6$, the CC contribution is attractive.

The repulsive CC is not limited to chlorides: one sees the same in $P_4$, $S_4$, $CS_2$, OCS, and indeed $Si_2H_6$, where it even leads to a repulsive inner-shell contribution overall. Some trends become clear along the homonuclear diatomic series: The first-row $A_2$ diatomics (Table 7, upper pane) have small attractive or repulsive CC terms, and a CV term that from $C_2$ to $F_2$ gradually decreases from decidedly attractive to mildly repulsive. From $P_2$ to $S_2$ to $Cl_2$, the CC term goes from strongly repulsive to nearly zero, while the attractive CV term decays in the same fashion and (T) provides much of the glue. For the $AX_n$ hydrides (Table 7, lower pane), the CV term takes on a repulsive value for $AlH_3$ but is otherwise attractive—it decays along the $CH_4$ to HF series. CC terms are small for the 1st row hydrides, and clearly repulsive for all 2nd row hydrides, tapering off along the series AlH3 to HCl. Indeed, the (T) contributions are repulsive for BH3 as well as for AlH3 and SiH4. For AlH3 and SiH4 (as well as, incidentally, for B2H6 and Si2H6) we likewise end up with repulsive overall inner-shell correlation corrections. The most straightforward explanation for all these observations entails the decreasing size of the inner-shell core from left to right within a row of the periodic table, as well as the 2s2p core in the 2nd row being bulkier to begin with.

**TABLE 7.** Breakdown of CCSD(T)/awCV5Z Inner-Shell Contributions to the TAE (kcal/mol) for the Homonuclear Diatomic Series and for the $AX_n$ hydrides

|       | CCSD  | CCSD(T) | (T)   | CC    | CV    | exclusion |
|-------|-------|---------|-------|-------|-------|-----------|
| $B_2$    | 0.49  | 0.77    | 0.28  | 0.01  | 0.62  | -0.14     |
| $C_2$    | 0.29  | 1.00    | 0.71  | -0.09 | 0.92  | -0.54     |
| $N_2$    | 0.44  | 0.76    | 0.32  | -0.15 | 0.71  | -0.11     |
| $O_2$    | 0.03  | 0.24    | 0.21  | -0.02 | 0.16  | -0.11     |
| $F_2$    | -0.34 | -0.09   | 0.25  | -0.05 | -0.10 | -0.19     |
| $P_2$    | -0.24 | 0.71    | 0.95  | -0.61 | 1.08  | -0.71     |
| $S_2$    | 0.02  | 0.50    | 0.48  | -0.25 | 0.59  | -0.32     |
| $Cl_2$   | -0.23 | 0.17    | 0.40  | -0.02 | 0.09  | -0.31     |
| $BH_3$   | 1.18  | 1.10    | -0.09 | 0.11  | 0.86  | 0.22      |
| $CH_4$   | 1.21  | 1.23    | 0.02  | 0.09  | 1.02  | 0.10      |
| $NH_3$   | 0.54  | 0.63    | 0.09  | 0.03  | 0.51  | 0.00      |
| $H_2O$   | 0.31  | 0.37    | 0.07  | 0.00  | 0.33  | -0.02     |
| HF    | 0.14  | 0.18    | 0.04  | -0.01 | 0.17  | -0.02     |
| $AlH_3$  | -0.51 | -0.79   | -0.28 | -1.01 | -0.27 | 0.77      |
| $SiH_4$  | -0.07 | -0.17   | -0.09 | -1.00 | 0.31  | 0.61      |
| $PH_3$   | 0.15  | 0.31    | 0.16  | -0.53 | 0.58  | 0.10      |
| $H_2S$   | 0.19  | 0.32    | 0.13  | -0.40 | 0.60  | -0.01     |
| HCl   | 0.12  | 0.19    | 0.07  | -0.22 | 0.37  | -0.03     |

In contrast, in (pseudo)hypervalent[81–85] compounds like HClO4, SF6, SO3, and ClF5 (Table 8), we note repulsive CV contributions paired with attractive CC contributions. (T) is also decidedly attractive. The inner-shell contribution is attractive for most of these systems, mostly because (T) and CC are outweighing the large negative CV contributions.

Can we make a concise general statement about the relative importance of CC and CV? The average of the CC/CV ratios for the whole sample will be skewed by a few cases where ΔTAE[CV] accidentally approaches zero. Instead, we can determine the averages of the absolute values of the CC and CV contributions, and compute their ratio $\sum_i|\Delta TAE[CC]_i|/\sum_i|\Delta TAE[CV]_i|$. It equals 0.185 for the entire W4-17 set, but increases to 0.473 for the 74 second-row molecules, compared to just 0.085 for the remaining first-row molecules. It is also worth mentioning that |ΔTAE[CC]/ΔTAE[CV]|≥1 for 19 molecules in the W4-17 set, 17 of them second-row. (The two first-row exceptions are FO2 and O3.) It is reasonable to assume that the relative importance of CC may increase still further in subsequent row of the Periodic Table.

**TABLE 8.** Breakdown of CCSD(T)/awCV5Z Inner-Shell Contributions to the TAE (kcal/mol) for a Selection of Pseudohypervalent Compounds

|  | CCSD | CCSD(T) | (T) | CC | CV | exclusion |
|---|---|---|---|---|---|---|
| $HClO_4$ | 0.13 | 0.99 | 0.85 | 1.53 | -1.49 | 0.10 |
| $PF_5$ | -0.22 | 0.28 | 0.50 | 1.10 | -2.01 | 0.69 |
| $SF_6$ | -0.89 | -0.17 | 0.72 | 2.17 | -3.56 | 0.49 |
| $ClF_5$ | -1.21 | -0.28 | 0.93 | 1.93 | -2.94 | -0.19 |
| $ClF_3$ | -0.59 | -0.01 | 0.58 | 0.93 | -1.30 | -0.23 |
| $PF_3$ | 0.23 | 0.62 | 0.39 | 0.52 | -0.54 | 0.24 |
| $HOClO_2$ | -0.06 | 0.67 | 0.73 | 1.04 | -1.00 | -0.10 |
| $ClOOCl$ | -0.62 | 0.16 | 0.78 | 0.04 | -0.09 | -0.56 |
| $ClO_3$ | -0.13 | 0.59 | 0.72 | 1.12 | -1.27 | 0.01 |
| $HOClO$ | -0.15 | 0.41 | 0.55 | 0.43 | -0.33 | -0.24 |
| $SO_2$ | 0.38 | 0.97 | 0.58 | 0.35 | 0.07 | -0.04 |
| $SO_3$ | 0.46 | 1.18 | 0.73 | 0.49 | -0.26 | 0.23 |

## General recommendations for thermochemistry

In W4 theory, we use what amounts to CCSD(T)/awCV{T,Q}Z. From Table 4, it appears that this is a smaller source of error than the valence correlation convergence or the post-CCSD(T) corrections. If we are to improve on this, the least expensive practical option is awCV{Q,5}Z: smaller-sized sequences are not adequate.

For a lower-cost solution that still yields thermochemically acceptable accuracy, we recommend combining CCSD-F12b/CVTZ-F12 with the (T) from a smaller basis set. This is, in fact, Karton's recommendation for the core-valence step in the original[14] W1-F12 and W2-F12 methods, as well as in the revised W2-F12 protocol specified in the W4-17 paper.[57] Specifically, he advocates determining the costly (T) component using a reduced cc-pwCVTZ(no f) basis set, and scaling this contribution by a constant factor of 1.1. By comparison with our best (T)/awCV{Q,5}Z, we find this relatively inexpensive basis set to work surprisingly well for this purpose. After scaling by 1.1, we find an RMSD of just 0.017 kcal/mol over the whole W4-17 set, compared to 0.049 for the original unscaled values. The scale factor that minimizes RMSD is found to be 1.127, for an RMSD of 0.014 kcal/mol. Combining this scaled small basis triples correction with CCSD-F12b/CVTZ-F12, we find here RMSD=0.10 kcal/mol, which can be reduced to just 0.06 kcal/mol through CV{D,T}Z-F12 extrapolation (with exponent 3.145 taken from Ref.[86]).

Performance for more economical methods was discussed in some detail by Ranasinghe et al.[62] (see Table 1 in that work) compared to {3,4}ZaPa-CV results, while our reference in the following is CCSD(T)awCV{Q,5}Z. Their findings (given in parentheses) for a different dataset only differ in detail from ours: CCSD(T)/MTsmall, which is used in W1 theory, has an RMSD of 0.14 (0.16) kcal/mol. Downgrading the electron correlation method to CCSD increases this statistic to 0.43 (0.39) kcal/mol. A further downgrade to MP2 degrades it to 0.77 (0.61) kcal/mol. (Ranasinghe et al. found 0.53 kcal/mol for MP4.) MP2 with the smaller G3LargeXP basis set,[1] as used in G4 theory, entails a similar RMSD of 0.75 (0.65) kcal/mol.

The error for the MP2(full)/awCV(T+d)Z–MP2(val)/aV(T+d)Z difference that is used in ccCA, 1.01 kcal/mol RMS, looks high but is actually comparable to the 0.95 kcal/mol found by Karton et al.[57] for the MP2/CVTZ based correction in W1X-n theory.[87] (It would appear that there is more error compensation going on in ccCA, G4, W1X-n, and similar methods than meets the eye.) Ranasinghe et al. actually found a much better RMSD of 0.27 kcal/mol for an ad hoc optimized 'core-valence DFT functional' of their own design.[62] This may actually be a more viable approach to low-cost core-valence calculations than MP2-based calculations.

## CONCLUSIONS

From the above benchmark study on the inner-shell contribution to total atomization energies, we can conclude the following:
• our best reference data for the W4-17 dataset should be reliable to 0.01 kcal/mol RMS overall, probably better than that for 1$^{st}$-row systems
• all three basis set families, aCVnZ, awCVnZ, and nZaPa-CV yield essentially equivalent results when extrapolating from the two largest basis sets
• for smaller values of n, nZaPa-CV performs best for core-core correlation and awCVnZ for core-valence correlation
• CCSD(T)/awCV{T,Q}Z, at 0.027 kcal/mol RMSD, is the best compromise between accuracy and computational cost for the inner-shell part of W4 and W4-F12; further improvement is only to be expected from awCV{Q,5}Z, but may not be worthwhile in light of remaining errors in the valence part;
• for lower-level calculations, CCSD-F12b/CV{D,T}Z-F12 combined with scaled (T)/pwCVTZ may offer a cost-effective alternative;
• MP2-level core-valence corrections are not recommended for accurate computational work; in the context of composite methods like G4 and ccCA, they benefit from error compensation;
• the widely accepted belief that inner-shell contributions are dominated by core-valence effects is largely correct for first-row compounds, but for second-row species, repulsive core-core interactions may become quite significant, particularly between adjacent 2$^{nd}$-row atoms. The average absolute core-core/core-valence ratio is 0.08 for the first-row species in W4-17, but 0.47 for the second-row subset.

## ACKNOWLEDGMENTS

This research was supported by the Israel Science Foundation (grant 1358/15) and by the Minerva Foundation, Munich, Germany, as well as by two internal Weizmann Institute funding sources: the Helen and Martin Kimmel Center for Molecular Design and a research grant from the estate of Emile Mimran. We thank Prof. Amir Karton (U. of Western Australia) for helpful discussions.

## SUPPORTING INFORMATION

Our best reference dataset for W4-17 at the CCSD(T)/awCV{Q,5}Z level (in Microsoft Excel format) and the aug-cc-pwCV(n+d)Z basis sets (n=T,Q; MOLPRO inline format) are freely available online in the FigShare data repository at http://doi.org/10.6084/m9.figshare.6154337, reference number 6154337.

## REFERENCES

(1) Curtiss, L. A.; Redfern, P. C.; Raghavachari, K. Gaussian-4 Theory. *J. Chem. Phys.* **2007**, *126* (8), 084108 DOI: 10.1063/1.2436888.
(2) Martin, J. M. L.; de Oliveira, G. Towards Standard Methods for Benchmark Quality Ab Initio Thermochemistry—W1 and W2 Theory. *J. Chem. Phys.* **1999**, *111* (5), 1843–1856 DOI: 10.1063/1.479454.
(3) Curtiss, L. A.; Redfern, P. C.; Raghavachari, K. Gn Theory. *Wiley Interdiscip. Rev. Comput. Mol. Sci.* **2011**, *1* (5), 810–825 DOI: 10.1002/wcms.59.
(4) Montgomery, J. A.; Frisch, M. J.; Ochterski, J. W.; Petersson, G. A. A Complete Basis Set Model Chemistry. VI. Use of Density Functional Geometries and Frequencies. *J. Chem. Phys.* **1999**, *110* (6), 2822 DOI: 10.1063/1.477924.
(5) Montgomery, J. A.; Frisch, M. J.; Ochterski, J. W.; Petersson, G. A. A Complete Basis Set Model Chemistry. VII. Use of the Minimum Population Localization Method. *J. Chem. Phys.* **2000**, *112* (15), 6532–6542 DOI: 10.1063/1.481224.

(6) Petersson, G. A. Complete Basis Set Models for Chemical Reactivity: From the Helium Atom to Enzyme Kinetics. In *Quantum-Mechanical Prediction of Thermochemical Data*; Cioslowski, J., Ed.; Kluwer Academic Publishers: Dordrecht, 2001; pp 99–130.

(7) DeYonker, N. J.; Wilson, B. R.; Pierpont, A. W.; Cundari, T. R.; Wilson, A. K. Towards the Intrinsic Error of the Correlation Consistent Composite Approach (CcCA). *Mol. Phys.* **2009**, *107* (8–12), 1107–1121 DOI: 10.1080/00268970902744359.

(8) DeYonker, N. J.; Cundari, T. R.; Wilson, A. K. The Correlation Consistent Composite Approach (CcCA): Efficient and Pan-Periodic Kinetics and Thermodynamics. In *Advances in the Theory of Atomic and Molecular Systems (Progress in Theoretical Chemistry and Physics, Vol. 19)*; Piecuch, P., Maruani, J., Delgado-Barrio, G., Wilson, S., Eds.; Progress in Theoretical Chemistry and Physics; Springer Netherlands: Dordrecht, 2009; Vol. 19, pp 197–224.

(9) DeYonker, N. J.; Cundari, T. R.; Wilson, A. K. The Correlation Consistent Composite Approach (CcCA): An Alternative to the Gaussian-n Methods. *J. Chem. Phys.* **2006**, *124* (11), 114104 DOI: 10.1063/1.2173988.

(10) Parthiban, S.; Martin, J. M. L. Assessment of W1 and W2 Theories for the Computation of Electron Affinities, Ionization Potentials, Heats of Formation, and Proton Affinities. *J. Chem. Phys.* **2001**, *114* (14), 6014–6029 DOI: 10.1063/1.1356014.

(11) Parthiban, S.; Martin, J. M. L. Fully Ab Initio Atomization Energy of Benzene via Weizmann-2 Theory. *J. Chem. Phys.* **2001**, *115* (5), 2051–2054 DOI: 10.1063/1.1385363.

(12) Karton, A.; Rabinovich, E.; Martin, J. M. L.; Ruscic, B. W4 Theory for Computational Thermochemistry: In Pursuit of Confident Sub-KJ/Mol Predictions. *J. Chem. Phys.* **2006**, *125* (14), 144108 DOI: 10.1063/1.2348881.

(13) Karton, A.; Taylor, P. R.; Martin, J. M. L. Basis Set Convergence of Post-CCSD Contributions to Molecular Atomization Energies. *J. Chem. Phys.* **2007**, *127* (6), 064104 DOI: 10.1063/1.2755751.

(14) Karton, A.; Martin, J. M. L. Explicitly Correlated Wn Theory: W1-F12 and W2-F12. *J. Chem. Phys.* **2012**, *136* (12), 124114 DOI: 10.1063/1.3697678.

(15) Chan, B.; Radom, L. W3X: A Cost-Effective Post-CCSD(T) Composite Procedure. *J. Chem. Theory Comput.* **2013**, *9* (11), 4769–4778 DOI: 10.1021/ct4005323.

(16) Chan, B.; Radom, L. W2X and W3X-L: Cost-Effective Approximations to W2 and W4 with KJ Mol −1 Accuracy. *J. Chem. Theory Comput.* **2015**, *11* (5), 2109–2119 DOI: 10.1021/acs.jctc.5b00135.

(17) Sylvetsky, N.; Peterson, K. A.; Karton, A.; Martin, J. M. L. Toward a W4-F12 Approach: Can Explicitly Correlated and Orbital-Based Ab Initio CCSD(T) Limits Be Reconciled? *J. Chem. Phys.* **2016**, *144* (21), 214101 DOI: http://dx.doi.org/10.1063/1.4952410.

(18) Feller, D.; Peterson, K. A.; Dixon, D. A. A Survey of Factors Contributing to Accurate Theoretical Predictions of Atomization Energies and Molecular Structures. *J. Chem. Phys.* **2008**, *129* (20), 204105 DOI: 10.1063/1.3008061.

(19) Li, S.; Hennigan, J. M.; Dixon, D. A.; Peterson, K. A. Accurate Thermochemistry for Transition Metal Oxide Clusters. *J. Phys. Chem. A* **2009**, *113* (27), 7861–7877 DOI: 10.1021/jp810182a.

(20) Bross, D. H.; Hill, J. G.; Werner, H.-J.; Peterson, K. A. Explicitly Correlated Composite Thermochemistry of Transition Metal Species. *J. Chem. Phys.* **2013**, *139* (9), 094302 DOI: 10.1063/1.4818725.

(21) Dixon, D.; Feller, D.; Peterson, K. A Practical Guide to Reliable First Principles Computational Thermochemistry Predictions Across the Periodic Table. *Annu. Rep. Comput. Chem.* **2012**, *8*, 1–28 DOI: 10.1016/B978-0-444-59440-2.00001-6.

(22) Feller, D.; Peterson, K. A.; Ruscic, B. Improved Accuracy Benchmarks of Small Molecules Using Correlation Consistent Basis Sets. *Theor. Chem. Acc.* **2013**, *133* (1), 1407 DOI: 10.1007/s00214-013-1407-z.

(23) East, A. L. L.; Allen, W. D. The Heat of Formation of NCO. *J. Chem. Phys.* **1993**, *99* (6), 4638–4650 DOI: 10.1063/1.466062.

(24) Allen, W. D.; East, A. L. L.; Császár, A. G. Ab Initio Anharmonic Vibrational Analyses of Non-Rigid Molecules. In *Structures and Conformations of Non-Rigid Molecules (NATO ASI Series 410)*; Laane, J., Dakkouri, M., Veken, B., Oberhammer, H., Eds.; Springer Netherlands: Dordrecht, 1993; pp 343–373.

(25) Wheeler, S. E.; Robertson, K. A.; Allen, W. D.; Schaefer; Bomble, Y. J.; Stanton, J. F. Thermochemistry of Key Soot Formation Intermediates: C 3 H 3 Isomers †. *J. Phys. Chem. A* **2007**, *111* (19), 3819–3830 DOI: 10.1021/jp0684630.

(26) Tajti, A.; Szalay, P. G.; Császár, A. G.; Kállay, M.; Gauss, J.; Valeev, E. F.; Flowers, B. A.; Vázquez, J.;

Stanton, J. F. HEAT: High Accuracy Extrapolated Ab Initio Thermochemistry. *J. Chem. Phys.* **2004**, *121* (23), 11599–11613 DOI: 10.1063/1.1811608.

(27) Harding, M. E.; Vázquez, J.; Ruscic, B.; Wilson, A. K.; Gauss, J.; Stanton, J. F. High-Accuracy Extrapolated Ab Initio Thermochemistry. III. Additional Improvements and Overview. *J. Chem. Phys.* **2008**, *128* (11), 114111 DOI: 10.1063/1.2835612.

(28) Harding, M. E.; Vázquez, J.; Gauss, J.; Stanton, J. F.; Kállay, M. Towards Highly Accurate Ab Initio Thermochemistry of Larger Systems: Benzene. *J. Chem. Phys.* **2011**, *135* (4), 044513 DOI: 10.1063/1.3609250.

(29) Karton, A.; Daon, S.; Martin, J. M. L. W4-11: A High-Confidence Benchmark Dataset for Computational Thermochemistry Derived from First-Principles W4 Data. *Chem. Phys. Lett.* **2011**, *510*, 165–178 DOI: 10.1016/j.cplett.2011.05.007.

(30) Ruscic, B.; Pinzon, R. E.; Morton, M. L.; von Laszevski, G.; Bittner, S. J.; Nijsure, S. G.; Amin, K. A.; Minkoff, M.; Wagner, A. F. Introduction to Active Thermochemical Tables: Several "Key" Enthalpies of Formation Revisited. *J. Phys. Chem. A* **2004**, *108* (45), 9979–9997 DOI: 10.1021/jp047912y.

(31) Ruscic, B.; Pinzon, R. E.; Laszewski, G. von; Kodeboyina, D.; Burcat, A.; Leahy, D.; Montoy, D.; Wagner, A. F. Active Thermochemical Tables: Thermochemistry for the 21st Century. *J. Phys. Conf. Ser.* **2005**, *16*, 561–570 DOI: 10.1088/1742-6596/16/1/078.

(32) Ruscic, B.; Pinzon, R. E.; Morton, M. L.; Srinivasan, N. K.; Su, M.-C.; Sutherland, J. W.; Michael, J. V. Active Thermochemical Tables: Accurate Enthalpy of Formation of Hydroperoxyl Radical, $HO_2$ †. *J. Phys. Chem. A* **2006**, *110* (21), 6592–6601 DOI: 10.1021/jp056311j.

(33) Stevens, W. R.; Ruscic, B.; Baer, T. Heats of Formation of $C_6H_5(\bullet)$, $C_6H_5(+)$, and $C_6H_5NO$ by Threshold Photoelectron Photoion Coincidence and Active Thermochemical Tables Analysis. *J. Phys. Chem. A* **2010**, *114* (50), 13134–13145 DOI: 10.1021/jp107561s.

(34) Ruscic, B. Active Thermochemical Tables: Water and Water Dimer. *J. Phys. Chem. A* **2013**, *117* (46), 11940–11953 DOI: 10.1021/jp403197t.

(35) Ruscic, B.; Feller, D.; Peterson, K. A. Active Thermochemical Tables: Dissociation Energies of Several Homonuclear First-Row Diatomics and Related Thermochemical Values. *Theor. Chem. Acc.* **2013**, *133* (1), 1415 DOI: 10.1007/s00214-013-1415-z.

(36) Ruscic, B. Active Thermochemical Tables: Sequential Bond Dissociation Enthalpies of Methane, Ethane, and Methanol and the Related Thermochemistry. *J. Phys. Chem. A* **2015**, *119* (28), 7810–7837 DOI: 10.1021/acs.jpca.5b01346.

(37) Fogueri, U. R.; Kozuch, S.; Karton, A.; Martin, J. M. L. A Simple DFT-Based Diagnostic for Nondynamical Correlation. *Theor. Chem. Acc.* **2012**, *132* (1), 1291 DOI: 10.1007/s00214-012-1291-y.

(38) Kesharwani, M. K.; Brauer, B.; Martin, J. M. L. Frequency and Zero-Point Vibrational Energy Scale Factors for Double-Hybrid Density Functionals (and Other Selected Methods): Can Anharmonic Force Fields Be Avoided? *J. Phys. Chem. A* **2015**, *119*, 1701–1714 DOI: 10.1021/jp508422u.

(39) Nesbet, R. K. Electronic Correlation in Atoms and Molecules. *Adv. Chem. Phys.* **1965**, *9*, 321–363 DOI: 10.1002/9780470143551.ch4.

(40) Bauschlicher, C. W.; Langhoff, S. R.; Taylor, P. R. Core–Core and Core–Valence Correlation. *J. Chem. Phys.* **1988**, *88* (4), 2540–2546 DOI: 10.1063/1.454032.

(41) Martin, J. M. L.; Taylor, P. R. Basis Set Convergence for Geometry and Harmonic Frequencies. Are h Functions Enough? *Chem. Phys. Lett.* **1994**, *225* (4–6), 473–479 DOI: 10.1016/0009-2614(94)87114-0.

(42) Martin, J. M. . On the Effect of Core Correlation on the Geometry and Harmonic Frequencies of Small Polyatomic Molecules. *Chem. Phys. Lett.* **1995**, *242* (3), 343–350 DOI: 10.1016/0009-2614(95)00747-R.

(43) Martin, J. M. L.; Parthiban, S. *Quantum-Mechanical Prediction of Thermochemical Data*; Cioslowski, J., Ed.; Understanding Chemical Reactivity; Kluwer Academic Publishers: Dordrecht, 2002; Vol. 22.

(44) Woon, D. E.; Dunning, T. H. Gaussian Basis Sets for Use in Correlated Molecular Calculations. V. Core-valence Basis Sets for Boron through Neon. *J. Chem. Phys.* **1995**, *103* (11), 4572–4585 DOI: 10.1063/1.470645.

(45) Peterson, K. A.; Dunning, T. H. Accurate Correlation Consistent Basis Sets for Molecular Core–Valence Correlation Effects: The Second Row Atoms Al–Ar, and the First Row Atoms B–Ne Revisited. *J. Chem. Phys.* **2002**, *117* (23), 10548–10560 DOI: 10.1063/1.1520138.

(46) Hill, J. G.; Mazumder, S.; Peterson, K. A. Correlation Consistent Basis Sets for Molecular Core-Valence Effects with Explicitly Correlated Wave Functions: The Atoms B-Ne and Al-Ar. *J. Chem. Phys.* **2010**, *132* (5), 054108 DOI: doi:10.1063/1.3308483.

(47) DeYonker, N. J.; Peterson, K. A.; Wilson, A. K. Systematically Convergent Correlation Consistent Basis Sets for Molecular Core?Valence Correlation Effects:? The Third-Row Atoms Gallium through Krypton ? *J. Phys. Chem. A* **2007**, *111* (44), 11383–11393 DOI: 10.1021/jp0747757.

(48) Balabanov, N. B.; Peterson, K. A. Systematically Convergent Basis Sets for Transition Metals. I. All-Electron Correlation Consistent Basis Sets for the 3d Elements Sc–Zn. *J. Chem. Phys.* **2005**, *123* (6), 064107 DOI: 10.1063/1.1998907.

(49) Peterson, K. A.; Figgen, D.; Dolg, M.; Stoll, H. Energy-Consistent Relativistic Pseudopotentials and Correlation Consistent Basis Sets for the 4d Elements Y–Pd. *J. Chem. Phys.* **2007**, *126* (12), 124101 DOI: 10.1063/1.2647019.

(50) Figgen, D.; Peterson, K. A.; Dolg, M.; Stoll, H. Energy-Consistent Pseudopotentials and Correlation Consistent Basis Sets for the 5d Elements Hf–Pt. *J. Chem. Phys.* **2009**, *130* (16), 164108 DOI: 10.1063/1.3119665.

(51) Peterson, K. A.; Puzzarini, C. Systematically Convergent Basis Sets for Transition Metals. II. Pseudopotential-Based Correlation Consistent Basis Sets for the Group 11 (Cu, Ag, Au) and 12 (Zn, Cd, Hg) Elements. *Theor. Chem. Acc.* **2005**, *114* (4–5), 283–296 DOI: 10.1007/s00214-005-0681-9.

(52) Hill, J. G.; Peterson, K. A. Correlation Consistent Basis Sets for Explicitly Correlated Wavefunctions: Pseudopotential-Based Basis Sets for the Post-d Main Group Elements Ga–Rn. *J. Chem. Phys.* **2014**, *141* (9), 094106 DOI: 10.1063/1.4893989.

(53) Iron, M. A.; Oren, M.; Martin, J. M. L. Alkali and Alkaline Earth Metal Compounds: Core—Valence Basis Sets and Importance of Subvalence Correlation. *Mol. Phys.* **2003**, *101* (9), 1345–1361 DOI: 10.1080/0026897031000094498.

(54) Hill, J. G.; Peterson, K. A. Correlation Consistent Basis Sets for Explicitly Correlated Wavefunctions: Valence and Core-Valence Basis Sets for Li, Be, Na, and Mg. *Phys. Chem. Chem. Phys.* **2010**, *12* (35), 10460–10468 DOI: 10.1039/c0cp00020e.

(55) Ranasinghe, D. S.; Frisch, M. J.; Petersson, G. A. Core-Core and Core-Valence Correlation Energy Atomic and Molecular Benchmarks for Li through Ar. *J. Chem. Phys.* **2015**, *143* (21), 214110 DOI: 10.1063/1.4935972.

(56) Ranasinghe, D. S.; Petersson, G. A. CCSD(T)/CBS Atomic and Molecular Benchmarks for H through Ar. *J. Chem. Phys.* **2013**, *138* (14), 144104 DOI: 10.1063/1.4798707.

(57) Karton, A.; Sylvetsky, N.; Martin, J. M. L. W4-17: A Diverse and High-Confidence Dataset of Atomization Energies for Benchmarking High-Level Electronic Structure Methods. *J. Comput. Chem.* **2017**, *38* (24), 2063–2075 DOI: 10.1002/jcc.24854.

(58) Trucks, G. W.; Pople, J. A.; Head-Gordon, M. A Fifth-Order Perturbation Comparison of Electron Correlation TheoriesA1 - Raghavachari,K. *Chem. Phys. Lett.* **1989**, *157*, 479–483.

(59) Watts, J. D.; Gauss, J.; Bartlett, R. J. Coupled-Cluster Methods with Noniterative Triple Excitations for Restricted Open-Shell Hartree–Fock and Other General Single Determinant Reference Functions. Energies and Analytical Gradients. *J. Chem. Phys.* **1993**, *98* (11), 8718–8733 DOI: 10.1063/1.464480.

(60) Martin, J. M. L.; Sundermann, A.; Fast, P. L.; Truhlar, D. G. Thermochemical Analysis of Core Correlation and Scalar Relativistic Effects on Molecular Atomization Energies. *J. Chem. Phys.* **2000**, *113* (4), 1348–1358 DOI: 10.1063/1.481960.

(61) Nicklass, A.; Peterson, K. A. Core-Valence Correlation Effects for Molecules Containing First-Row Atoms. Accurate Results Using Effective Core Polarization Potentials. *Theor. Chem. Acc.* **1998**, *100* (1–4), 103–111 DOI: 10.1007/s002140050370.

(62) Ranasinghe, D. S.; Frisch, M. J.; Petersson, G. A. A Density Functional for Core-Valence Correlation Energy. *J. Chem. Phys.* **2015**, *143* (21), 214111 DOI: 10.1063/1.4935973.

(63) Werner, H.-J.; Knowles, P. J.; Knizia, G.; Manby, F. R.; Schütz, M.; Celani, P.; Korona, T.; Lindh, R.; Mitrushenkov, A.; Rauhut, G.; Shamasundar, K. R.; Adler, T. B.; Amos, R. D.; Bernhardsson, A.; Berning, A.; Cooper, D. L.; Deegan, M. J. O.; Dobbyn, A. J.; Eckert, F.; Goll, E.; Hampel, C.; Hesselman, A.; Hetzer, G.; Hrenar, T.; Jansen, G.; Köppl, C.; Liu, Y.; Lloyd, A. W.; Mata, R. A.; May, A. J.; McNicholas, S. J.; Meyer, W.; Mura, M. E.; Nicklass, A.; O'Neill, D. P.; Palmieri, P.; Peng, D.; Pflüger, K.; Pitzer, R. M.; Reiher, M.; Shiozaki, T.; Stoll, H.; Stone, A. J.; Tarroni, R.; Thorsteinsson, T.; Wang, M. MOLPRO, Version 2015.1, a Package of Ab Initio Programs. University of Cardiff Chemistry Consultants (UC3): Cardiff, Wales, UK 2015.

(64) Werner, H.-J.; Knowles, P. J.; Knizia, G.; Manby, F. R.; Schütz, M. Molpro: A General-Purpose Quantum Chemistry Program Package. *Wiley Interdiscip. Rev. Comput. Mol. Sci.* **2012**, *2* (2), 242–253 DOI:

(65) Kong, L.; Bischoff, F. A.; Valeev, E. F. Explicitly Correlated R12/F12 Methods for Electronic Structure. *Chem. Rev.* **2012**, *112* (1), 75–107 DOI: 10.1021/cr200204r.
(66) Hättig, C.; Klopper, W.; Köhn, A.; Tew, D. P. Explicitly Correlated Electrons in Molecules. *Chem. Rev.* **2012**, *112* (1), 4–74 DOI: 10.1021/cr200168z.
(67) Knizia, G.; Adler, T. B.; Werner, H.-J. Simplified CCSD(T)-F12 Methods: Theory and Benchmarks. *J. Chem. Phys.* **2009**, *130* (5), 054104 DOI: 10.1063/1.3054300.
(68) Adler, T. B.; Knizia, G.; Werner, H.-J. A Simple and Efficient CCSD(T)-F12 Approximation. *J. Chem. Phys.* **2007**, *127* (22), 221106 DOI: 10.1063/1.2817618.
(69) Noga, J.; Šimunek, J. On the One-Particle Basis Set Relaxation in R12 Based Theories. *Chem. Phys.* **2009**, *356* (1–3), 1–6 DOI: 10.1016/j.chemphys.2008.10.012.
(70) Ten-no, S. Initiation of Explicitly Correlated Slater-Type Geminal Theory. *Chem. Phys. Lett.* **2004**, *398* (1–3), 56–61 DOI: 10.1016/j.cplett.2004.09.041.
(71) Weigend, F. A Fully Direct RI-HF Algorithm: Implementation, Optimised Auxiliary Basis Sets, Demonstration of Accuracy and Efficiency. *Phys. Chem. Chem. Phys.* **2002**, *4* (18), 4285–4291 DOI: 10.1039/b204199p.
(72) Hättig, C. Optimization of Auxiliary Basis Sets for RI-MP2 and RI-CC2 Calculations: Core–Valence and Quintuple-ζ Basis Sets for H to Ar and QZVPP Basis Sets for Li to Kr. *Phys. Chem. Chem. Phys.* **2005**, *7* (1), 59–66 DOI: 10.1039/b415208e.
(73) Valeev, E. F. Improving on the Resolution of the Identity in Linear R12 Ab Initio Theories. *Chem. Phys. Lett.* **2004**, *395* (4–6), 190–195 DOI: 10.1016/j.cplett.2004.07.061.
(74) Yousaf, K. E.; Peterson, K. A. Optimized Auxiliary Basis Sets for Explicitly Correlated Methods. *J. Chem. Phys.* **2008**, *129* (18), 184108 DOI: 10.1063/1.3009271.
(75) Brauer, B.; Kesharwani, M. K.; Kozuch, S.; Martin, J. M. L. The S66x8 Benchmark for Noncovalent Interactions Revisited: Explicitly Correlated Ab Initio Methods and Density Functional Theory. *Phys. Chem. Chem. Phys.* **2016**, *18* (31), 20905–20925 DOI: 10.1039/C6CP00688D.
(76) Sinanoğlu, O. Electron Correlation in Atoms and Molecules. *Adv. Chem. Phys.* **1969**, *14*, 237–282 DOI: 10.1002/9780470143599.ch8.
(77) Yockel, S.; Wilson, A. K. Core-Valence Correlation Consistent Basis Sets for Second-Row Atoms (Al–Ar) Revisited. *Theor. Chem. Acc.* **2008**, *120* (1–3), 119–131 DOI: 10.1007/s00214-007-0309-3.
(78) Marchetti, O.; Werner, H.-J. Accurate Calculations of Intermolecular Interaction Energies Using Explicitly Correlated Wave Functions. *Phys. Chem. Chem. Phys.* **2008**, *10* (23), 3400–3409 DOI: 10.1039/b804334e.
(79) Peterson, K. A.; Kesharwani, M. K.; Martin, J. M. L. The Cc-PV5Z-F12 Basis Set: Reaching the Basis Set Limit in Explicitly Correlated Calculations. *Mol. Phys.* **2015**, *113* (13–14), 1551–1558 DOI: 10.1080/00268976.2014.985755.
(80) Mann, J. B. *Atomic Structure Calculations. II. Hartree-Fock Wavefunctions and Radial Expectation Values: Hydrogen to Lawrencium (LANL Report LA-3691)*; Los Alamos, NM (United States), 1968.
(81) Magnusson, E. Hypercoordinate Molecules of Second-Row Elements: D Functions or d Orbitals? *J. Am. Chem. Soc.* **1990**, *112* (22), 7940–7951 DOI: 10.1021/ja00178a014.
(82) Reed, A. E.; Schleyer, P. v. R. Chemical Bonding in Hypervalent Molecules. The Dominance of Ionic Bonding and Negative Hyperconjugation over d-Orbital Participation. *J. Am. Chem. Soc.* **1990**, *112* (4), 1434–1445 DOI: 10.1021/ja00160a022.
(83) Cioslowski, J.; Mixon, S. T. Rigorous Interpretation of Electronic Wave Functions. 2. Electronic Structures of Selected Phosphorus, Sulfur, and Chlorine Fluorides and Oxides. *Inorg. Chem.* **1993**, *32* (15), 3209–3216 DOI: 10.1021/ic00067a004.
(84) Martin, J. M. L. Heats of Formation of Perchloric Acid, HClO4, and Perchloric Anhydride, Cl2O7. Probing the Limits of W1 and W2 Theory. *J. Mol. Struct. THEOCHEM* **2006**, *771* (1–3), 19–26 DOI: 10.1016/j.theochem.2006.03.035.
(85) Crabtree, R. H. Hypervalency, Secondary Bonding and Hydrogen Bonding: Siblings under the Skin. *Chem. Soc. Rev.* **2017**, *46* (6), 1720–1729 DOI: 10.1039/C6CS00688D.
(86) Hill, J. G.; Peterson, K. A.; Knizia, G.; Werner, H.-J. Extrapolating MP2 and CCSD Explicitly Correlated Correlation Energies to the Complete Basis Set Limit with First and Second Row Correlation Consistent Basis Sets. *J. Chem. Phys.* **2009**, *131* (19), 194105 DOI: 10.1063/1.3265857.
(87) Chan, B.; Radom, L. W1X-1 and W1X-2: W1-Quality Accuracy with an Order of Magnitude Reduction in Computational Cost. *J. Chem. Theory Comput.* **2012**, *8* (11), 4259–4269 DOI: 10.1021/ct300632p.

16